\documentclass[aip,reprint]{revtex4-2}

\usepackage{graphicx}
\usepackage{color}
\usepackage{amsmath,amssymb}
\usepackage{bm}
\usepackage{upgreek}
\usepackage{hyperref}
\usepackage{physics}
\hypersetup
{
	colorlinks=true,
	allcolors=blue,
}

\newcommand{\red}[1] {{\color{black}{#1}}}
\newcommand{\x}[1] {\text{#1}}

\begin{document}

\title{All-carbon approach to inducing electrical and optical anisotropy in graphene}

\author{Aleandro Antidormi}
\email{aleandro.antidormi@icn2.cat}
\affiliation{Catalan Institute of Nanoscience and Nanotechnology (ICN2), CSIC and BIST, Campus UAB, Bellaterra, 08193 Barcelona, Spain}

\author{Aron W. Cummings}
\email{aron.cummings@icn2.cat}
\affiliation{Catalan Institute of Nanoscience and Nanotechnology (ICN2), CSIC and BIST,
Campus UAB, Bellaterra, 08193 Barcelona, Spain}

\begin{abstract}
Owing to its array of unique properties, graphene is a promising material for a wide variety of applications. Being two-dimensional, the properties of graphene are also easily tuned via proximity to other materials. In this work, we investigate the possibility of inducing electrical and optical anisotropy in graphene by interfacing it with other anisotropic carbon systems, including nanoporous graphene and arrays of graphene nanoribbons. We find that such materials do indeed induce such anisotropy in graphene, while also preserving the unique properties offered by graphene's Dirac band structure, namely its superior charge transport and long-wavelength optical absorption. The optical anisotropy makes such heterostructures interesting for their use in applications related to long-wavelength polarimetry, while the electrical anisotropy may be valuable for enhancing the performance of graphene photothermoelectric detectors.
\end{abstract}

\maketitle

Graphene possesses unique properties with potential for a variety of applications \cite{Ferrari2015}. 
With respect to its optical properties, the absence of a band gap and its linear Dirac band structure are responsible for its uniform light absorption over a broad range of frequencies, from the optical to the THz regime \cite{Dawlaty2008}.
Moreover, graphene is fully compatible with standard silicon photonics and exhibits tunable electro-absorption and electro-refraction with fast electron dynamics and a small Fermi surface.
These factors are motivating the development of graphene-based photonics with applications in optical data communications \cite{Romagnoli2018, Schuler2016, Miseikis2020, Marconi2021}, THz technologies \cite{Cai2014, Koppens2019}, and plasmonics \cite{Koppens2011, Grigorenko2012}.

Another fundamental feature of graphene is the tunability of its physical properties, which may be achieved electrostatically, chemically, or via proximity to other materials. As an example, strong spin-orbit coupling can be induced in graphene by interfacing it with transition metal dichalcogenides, leading to phenomena such as large spin relaxation anisotropy and spin-charge conversion \cite{Cummings2017, Benitez2020}. Meanwhile, graphene's Dirac cone and superior charge transport are maintained in these systems, allowing the transfer of spin over long distances.

In the ongoing search for new functionalities of graphene, one strategy is to alter its features by imposing long range periodicity, called a superlattice, on top of its underlying crystal structure. This can be done with top-down fabrication methods, such as lithography of the dielectric substrate \cite{Forsythe2018, Li2021}, with feature sizes on the order of tens of nanometers. Superlattices can also be induced by layering graphene with hexagonal boron nitride or another layer of graphene, leading to exotic electronic properties \cite{Yankowitz2012, Cao2018}. In these systems, the periodicity of the Moir\'e superlattice is tunable by varying the twist angle between the two layers.

Recently, superlattices with periodicity $\approx$1 nm have been realized in graphene. By using directed
reactions among self-organized molecular precursors, it is possible to fabricate graphene containing a periodic array of nanoscale holes \cite{Moreno2018}. Called nanoporous graphene (NPG), this material has a large band gap and is expected to exhibit anisotropic optical and electrical properties, making it intriguing for devices. It also represents a new approach for customizing graphene, by using chemistry to design nanoscale superlattices from the bottom up. Similar techniques can also grow arrays of graphene nanoribbons (GNRs) \cite{Moreno2018b}.

In this paper, we explore the possibility of tuning the properties of graphene by interfacing it with NPG or arrays of GNRs. The goal is to combine one of the interesting properties of these materials, namely their anisotropy, with the unique properties offered by graphene's Dirac band structure, namely its superior charge transport and long-wavelength optical absorption \red{into the mid-IR and beyond}. Using numerical simulations, we show that NPG and GNRs can indeed induce anisotropy in graphene, both in its optical absorption and in its electrical transport.

With respect to optical absorption, this opens up graphene for its potential use in IR polarimetry. \red{Polarimetry is the use of the polarization of detected light to analyze images better than can be done only with intensity and wavelength. In the IR region, analysis of the polarity of detected light can be used, e.g., for the detection of land mines \cite{Tyo2006, Gurton2012}, to conduct tissue analysis \cite{Demos1997, Guo2004, Ghosh2011}, or to study astrophysical phenomena \cite{Aitken2004, Chrysostomou2007}}. 

Meanwhile, anisotropic charge transport may be valuable for enhancing the efficiency of graphene photodetectors based on the photothermoelectric effect \cite{Schuler2016, Miseikis2020, Marconi2021, Cai2014, Koppens2019}. A recent extensive analysis indicates that suppressing the electronic thermal conductivity in the transverse direction of such photodetectors, while maintaining it in the longitudinal direction, would enhance their performance \cite{Antidormi2021}.

\begin{figure}[tbh]
\centering
\includegraphics[width=8.5cm]{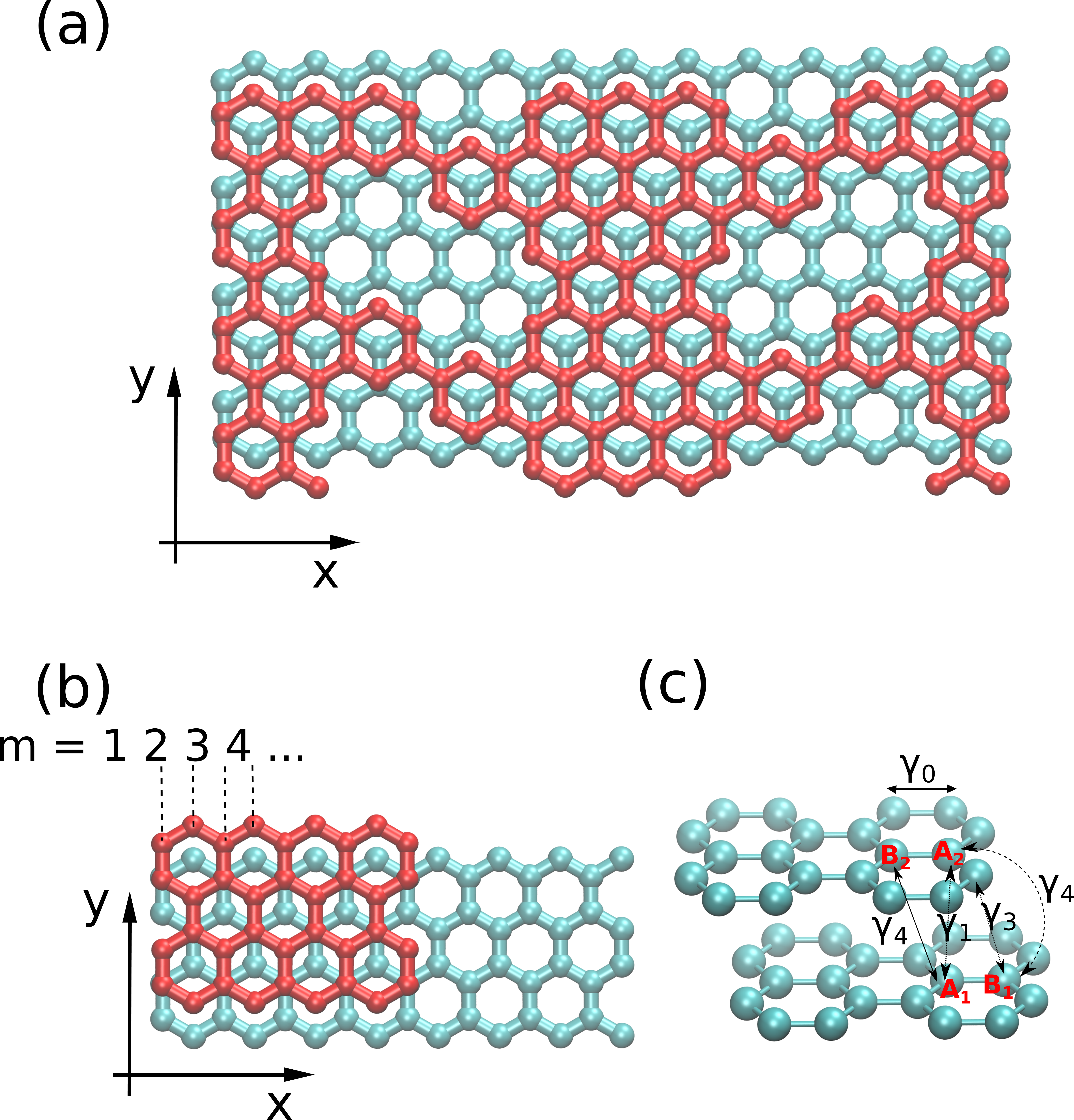}
\caption{\red{Two unit cells} of (a) the graphene/NPG heterostructure and (b) a graphene/9-aGNR heterostructure, both with AB stacking. Light blue atoms belong to graphene, and dark red atoms belong to the NPG or GNR. (c) The hopping parameters used in the tight-binding model of Eq.\ \eqref{eq_hamiltonian}.}
\label{fig_struct_npg}
\end{figure}

The system we primarily focus on in this work is shown in Fig.\ \ref{fig_struct_npg}. Panel (a) shows \red{two unit cells (along the $y$ axis)} of a graphene/NPG heterostructure in the lowest-energy AB stacking configuration. The NPG is the same as that synthesized in Ref.\ \citenum{Moreno2018} and may be viewed as an array of GNRs oriented along the $y$ axis and linked together via a single carbon-carbon bond. Panel (c) shows the hopping parameters that are used in our tight-binding model of this system, with the Hamiltonian given by
\begin{align}
\hat{H}  = &-\gamma_0 \sum_{\left<ij\right>, m} \left(\hat{a}^{\dagger}_{m,i} \hat{b}_{m,j} + \x{h.c.} \right) \nonumber \\
 &-\gamma_1 \sum_{j} \left(\hat{a}^{\dagger}_{1,j} \hat{a}_{2,j} + \x{h.c.} \right) \nonumber \\
 &-\gamma_3 \sum_{j} \left(\hat{b}^{\dagger}_{1,j} \hat{b}_{2,j} + \x{h.c.} \right) \nonumber \\
 &-\gamma_4 \sum_{j} \left(\hat{a}^{\dagger}_{1,j} \hat{b}_{2,j}  +  \hat{a}^{\dagger}_{2,j} \hat{b}_{1,j} + \x{h.c.} \right),
 \label{eq_hamiltonian}
\end{align}
where $\hat{a}^{\dagger}_{m,j}$ and $\hat{b}^{\dagger}_{m,j}$ ($\hat{a}_{m,j}$ and $\hat{b}_{m,j}$) are the creation (annihilation) operators of the $p_z$ orbital on sublattice A or B respectively, at lattice site $j$ of layer $m= 1,2$.
The parameter $\gamma_0 =2.9$ eV is the in-plane hopping energy between nearest neighbors, while $\gamma_1 = 0.4$ eV, $\gamma_3 = 0.3$ eV, and $\gamma_4 = 0.04$ eV are the interlayer hoppings, as depicted in Fig.\ \ref{fig_struct_npg}(c). The values of these parameters are chosen to be those of bilayer graphene \cite{CastroNeto2009}. 

\begin{figure*}[tbh]
\centering
\includegraphics[width=17cm]{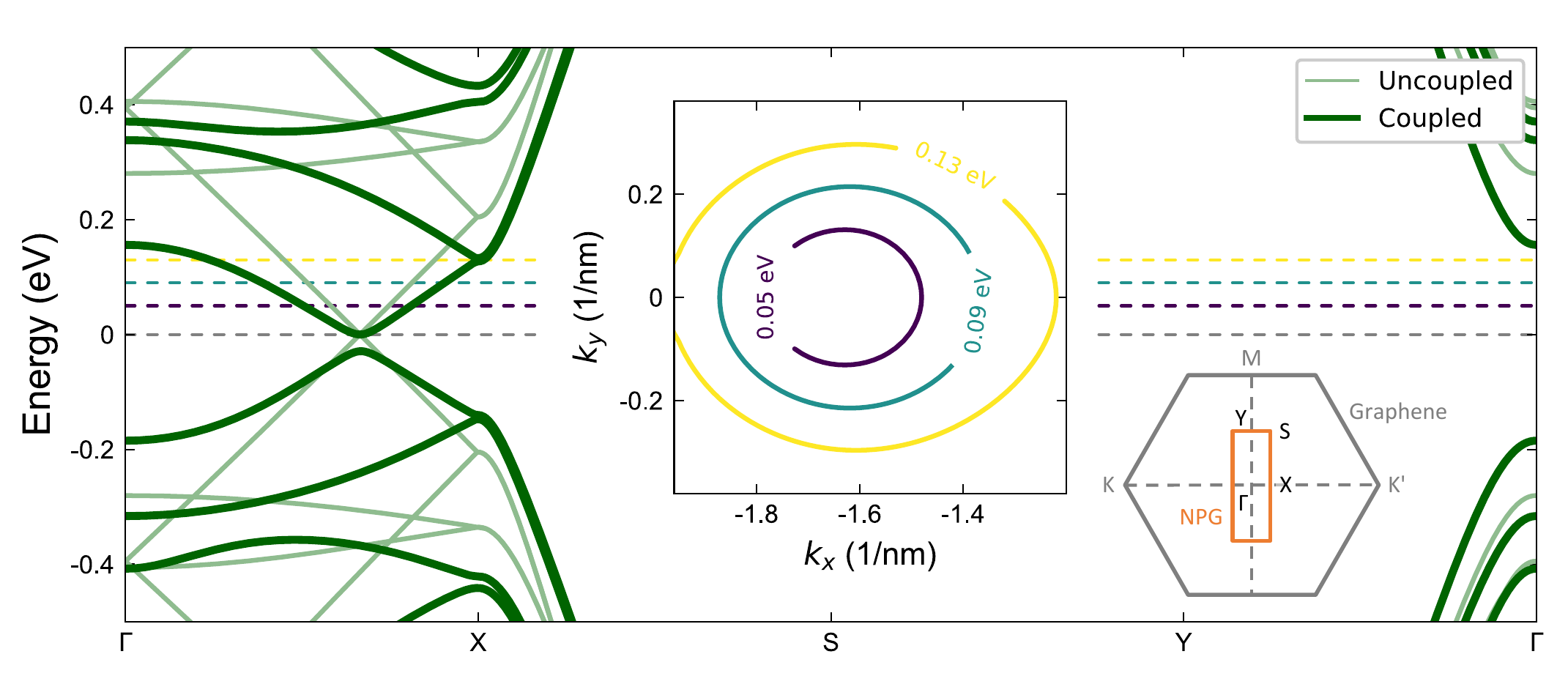}
\caption{Band structure of the graphene/NPG heterostructure of Fig.\ \ref{fig_struct_npg}(a). The thin faint lines show the band structure when the two layers are uncoupled, while the thicker darker lines are when the interlayer coupling is turned on. The middle inset shows the Fermi surface of the \red{graphene-like} Dirac bands at a few energies, highlighting the anisotropy induced by hybridization with the NPG. The lower right inset shows the Brillouin zone of the isolated graphene layer and that of the graphene/NPG heterostructure.}
\label{fig_band_structure}
\end{figure*}

From this Hamiltonian we can compute the electronic band structure, the optical absorption, and the transport properties of the NPG/graphene heterostructure. We begin with the band structure, shown in Fig.\ \ref{fig_band_structure}. The thinner, fainter lines show the band structure of the two layers when they are uncoupled ($\gamma_1 = \gamma_3 = \gamma_4 = 0$), i.e., they are the superposition of the band structures of the individual graphene and NPG layers. Owing to the larger supercell, the graphene Dirac cones have been folded from K/K' to positions along the $\Gamma$-X line. Meanwhile, the minimum (maximum) of the conduction (valence) band of the NPG remains at $\Gamma$, with a band gap of 0.56 eV.

The thicker, bolder lines show the band structure when we turn on the interlayer coupling. \red{Owing to hybridization between the layers, the bands no longer strictly belong to states in the graphene or NPG layer. Henceforth, we refer to the bands around $\Gamma$ as ``NPG-like'' and those that belong to the graphene Dirac cone along $\Gamma$-X as ``graphene-like.''} In the presence of interlayer coupling, the band gap of the \red{NPG-like bands} shrinks to 0.34 eV. The \red{graphene-like} bands remain within this gap but also exhibit a small gap opening of 32 meV. More striking, however, is a strong renormalization of the Fermi velocity of the \red{graphene-like} bands. This can be seen in the middle inset, where we plot the Fermi contour at a few energies near the charge neutrality point. The elliptical shape of the Fermi surface indicates the anisotropy induced in the \red{graphene-like} bands due to hybridization with the NPG. The Fermi velocity along the $x$-direction is suppressed with respect to that along the $y$ direction, with the ratio $v_{\x{F},x} / v_{\x{F},y} \approx 1.2$ at a Fermi energy of 0.1 eV, and $v_{\x{F},x} / v_{\x{F},y} \approx 1.4$ near the bottom of the Dirac cone.

The anisotropy induced in the band structure of the graphene by the NPG is expected to affect its optical and electronic properties. We first focus on the optical absorption by computing the optical conductivity of the graphene/NPG heterostructure. In the electric dipole approximation, the optical conductivity is given by \cite{Davies1997}
\begin{gather}
\sigma(\omega) = \frac{2 \pi e^2}{m_0^2 \omega \Omega} \sum_{i,j} \left| \bra{j} \mathbf{e} \cdot \mathbf{\hat{p}} \ket{i} \right|^2 \nonumber \\
\times \left[ f(E_i) - f(E_j) \right] \delta\left(E_j - E_i - \hbar \omega \right),
\label{eq_opt_cond}
\end{gather}
where $\omega$ is the optical frequency, $m_0$ is the free electron mass, $\mathbf{\hat{p}}$ is the momentum operator, $f(E)$ is the Fermi distribution function, and $\Omega$ is the volume of the unit cell. The unit vector $\mathbf{e}$ denotes the polarization of the electric field, which is real for linear polarization.
Equation \eqref{eq_opt_cond} expresses the conductivity as the sum of all vertical transitions between states $\ket{i}$ and $\ket{j}$, with the $\delta$ term imposing conservation of energy.  
The translational invariance of the system allows us to use the Bloch eigenstates of $\hat{H}$, $\ket{\mathbf{k}}$, and to use the relation
\begin{equation}
\bra{j} \mathbf{e} \cdot \mathbf{\hat{p}} \ket{i} = \frac{i m_0}{\hbar} \bra{j} \mathbf{e} \cdot \frac{\partial \hat{H}}{\partial  \mathbf{k}} \ket{i},
\end{equation}
which involves the derivative of the Hamiltonian with respect to the wavevector $\mathbf{k}$.
The sum in Eq.\ \eqref{eq_opt_cond} is then performed via numerical integration over the Brillouin zone with 701 $\times$ 701 points, while the $\delta$-function is approximated as a Lorentzian with a broadening of 10 meV. Here we ignore excitonic effects, as they are not present in the linear portion of the graphene spectrum \cite{Yang2009}, and the first excitonic absorption peak of NPG is predicted to be above 1 eV \cite{Singh2020}, beyond our range of interest.

\begin{figure}[tbh]
\centering
\includegraphics[width=8.5cm]{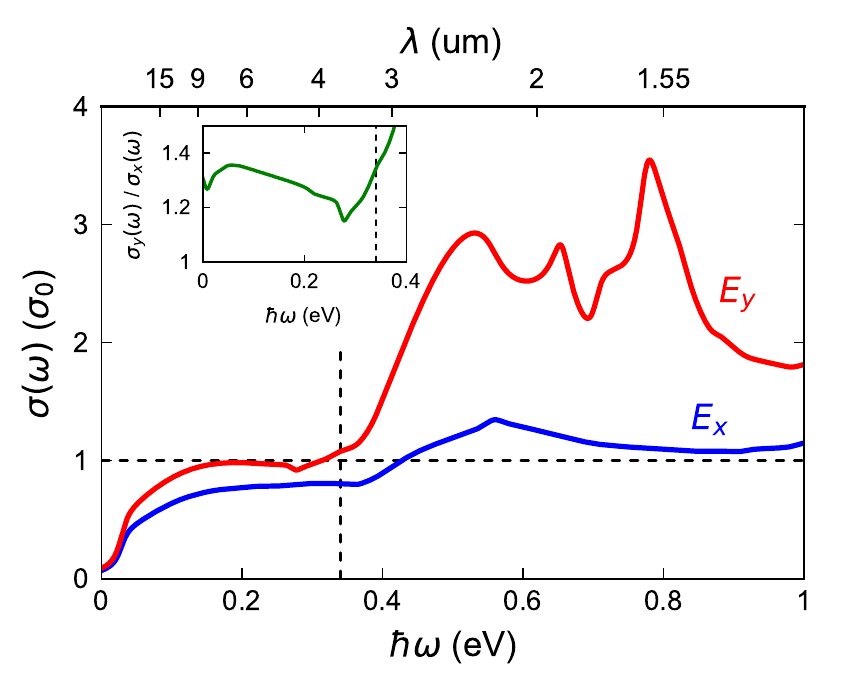}
\caption{Optical conductivity of the graphene/NPG heterostructure for light linearly polarized along the $x$ and $y$ axes. The vertical dashed line indicates the onset of absorption within the \red{NPG-like bands at $\Gamma$}. The inset shows the optical anisotropy, given by the ratio $\sigma_y(\omega) / \sigma_x(\omega)$.}
\label{fig_opt_cond}
\end{figure}

The optical conductivity of the NPG/graphene heterostructure is shown in Fig.\ \ref{fig_opt_cond} as a function of photon energy\red{/wavelength} at $T = 300$ K. Its value is normalized with respect to the ``universal'' optical conductivity of graphene, $\sigma_0 = \pi e^2/(2h)$ \cite{Gusynin2006}. The curves labeled $E_x$ and $E_y$ correspond to linearly-polarized light aligned along the $x$ and $y$ axes, respectively. The optical conductivity is clearly anisotropic over the entire energy range, with larger absorption for optical fields aligned along $y$. This anisotropy is largest for $\hbar\omega > 0.34$ eV, \red{when optical absorption begins to occur in non-graphene-like bands}. 

For photon energies smaller than 0.34 eV, highlighted by the vertical dashed line, only \red{graphene-like} states are present and the resulting conductivity involves only optical transitions \red{within these bands}. In this case, we also observe that $\sigma_y(\omega)$ is larger than $\sigma_x(\omega)$ as a consequence of the anisotropy of the \red{graphene-like} bands induced by NPG. As shown in the inset, the magnitude of this anisotropy is $\sigma_y(\omega)/\sigma_x(\omega) \approx 1.2-1.4$. This anisotropy is similar to that in photodetectors based on aligned arrays of carbon nanotubes \cite{He2013}, indicating that NPG is indeed effective at inducing optical anisotropy in graphene.

The optical anisotropy observed for small energies is associated with anisotropic modification of the graphene Fermi velocity induced by the NPG. We expect a similar impact on charge transport \red{at these energies}.
To examine this, we compute the electrical conductivity using a real-space order-$N$ wave packet propagation method \cite{Fan2020}.
The key quantity of this method is the energy- and time-dependent mean-square displacement of the wave packet,
\begin{equation}
\Delta X^2(E,t) =  \frac{ \Trace{ \left[ \delta(E-\hat{H}) \left|\hat{X}(t) - \hat{X}(0)\right|^2 \right] } }{ \rho(E)},
\end{equation}
where $\rho(E) = \Trace [ \delta(E - \hat{H}) ]$ is the density of states and $\hat{X}$ is the position operator along the $x$ axis. From this we calculate the time-dependent diffusion coefficient $D_{xx}(E,t) = \frac{1}{2}\frac{\partial }{\partial t} \Delta X^2(E,t)$ and its long-time limit $\widetilde{D}_{xx}(E)$. The latter enters in the resulting electronic conductivity, $\sigma_{xx}(E) = e^2 \rho(E) \widetilde{D}_{xx}(E)$. By separately computing the evolution of the mean-square displacement along $x$ and $y$ directions, we can evaluate the conductivities $\sigma_{xx}(E)$ and $\sigma_{yy}(E)$. 

We assume transport is dominated by charged impurity scattering, which is modeled as a random distribution of Gaussian electrostatic impurities \cite{Adam2009}. The electrostatic potential at each atomic site $i$ is then given by $\epsilon_i = \sum_j V_j \exp(-|\vec{r}_i-\vec{r}_j|^2 / 2\xi^2)$, where $\vec{r}_i$ is the position of each carbon atom, $\vec{r}_j$ is the position of each impurity, $\xi$ is the width of each impurity, and the height $V_j$ of each impurity is randomly distributed in $[-V,V]$. Here we use $V = 2.8$ eV, $\xi = \sqrt{3}a$, and an impurity density of $0.1\%$. This choice of parameters leads to a charge mobility of around 2000 cm$^2$/Vs, similar to experiments \cite{Chen2008}. 

All calculations have been performed on a graphene/NPG heterostructure with $22 \times 10^6$ atoms, consisting of $672\times168$ replicas of the unit cell shown in Fig.\ \ref{fig_struct_npg}(a). The conductivity has been averaged over ten randomly chosen initial wave packets, and the calculation of the mean-square displacement employed an efficient Chebyshev polynomial expansion with 5000 moments, corresponding to an energy resolution of $\sim$10 meV.

\begin{figure}[htbp]
\centering
\includegraphics[width=8.5cm]{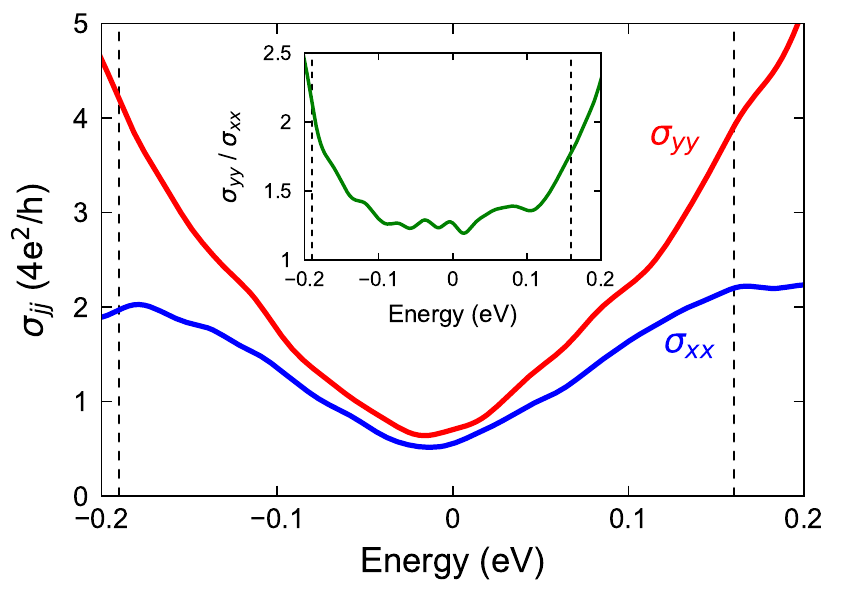}
\caption{Longitudinal electrical conductivity of the graphene/NPG heterostructure along the $x$ and $y$ directions. The vertical dashed lines indicate the onset of \red{the NPG-like bands at $\Gamma$ (see Fig.\ \ref{fig_band_structure})}. The inset shows the anisotropy of charge transport, given by $\sigma_{yy}/\sigma_{xx}$.}
\label{fig_cond}
\end{figure}

The electronic conductivities of the NPG/graphene heterostructure along the $x$ and $y$ directions are shown in Fig.\ \ref{fig_cond}. At low energies transport is only within the \red{graphene-like bands}, with the vertical dashed lines indicating the onset of transport in the \red{NPG-like bands at $\Gamma$ (see Fig.\ \ref{fig_band_structure})}.
As with the optical conductivity, the electrical conductivity along $y$ is larger than that along $x$. The resulting anisotropy is shown in the inset and takes on values of $\sigma_{yy}/\sigma_{xx} \approx 1.2-1.4$, the same as the optical conductivity. Upon the onset of charge transport in the \red{NPG-like bands}, the anisotropy grows dramatically as a result of saturation of $\sigma_{xx}$.

We have shown that interfacing graphene with NPG induces anisotropy in its optical and electrical conductivities, which is a result of the anisotropic renormalization of the graphene Dirac cones. We now examine this behavior for graphene interfaced with other carbon-based anisotropic 2D systems. Specifically, we consider two other configurations of NPG that were studied in Ref.\ \citenum{Gaetano2019}. These structures are the same as that in Fig.\ \ref{fig_struct_npg} except that, instead of a single carbon-carbon bond connecting the GNRs, a benzene ring serves as the bridge. It was found that the connection of this benzene bridge (meta vs.\ para) had a notable impact on the charge transport anisotropy in the NPG. Here we see if this difference carries over to the anisotropy induced in graphene. \red{To test the impact of the NPG bandgap on the optical anisotropy induced in graphene, we have also created two new NPG samples that have the same pore structure as that in Fig.\ \ref{fig_struct_npg}(a), but with wider regions between the pores. We call these structures NPG-8 and NPG-10, as they are 8 and 10 carbon rings wide, respectively, at the widest point.}
 
 We also investigate heterostructures of graphene with periodic arrays of armchair graphene nanoribbons (aGNRs). We consider the collection of $m$-aGNRs, with $m = \{3p,3p+1\}$, where $m$ is the number of atoms across the width of the ribbon and $p>0$ is an integer. These aGNRs are semiconducting, with their band gap inversely proportional their width \cite{Nakada1996}. A unit cell of graphene interfaced with a 9-aGNR is shown in Fig.\ \ref{fig_struct_npg}(b).

\begin{figure}[tbh]
\centering
\includegraphics[width=8.5cm]{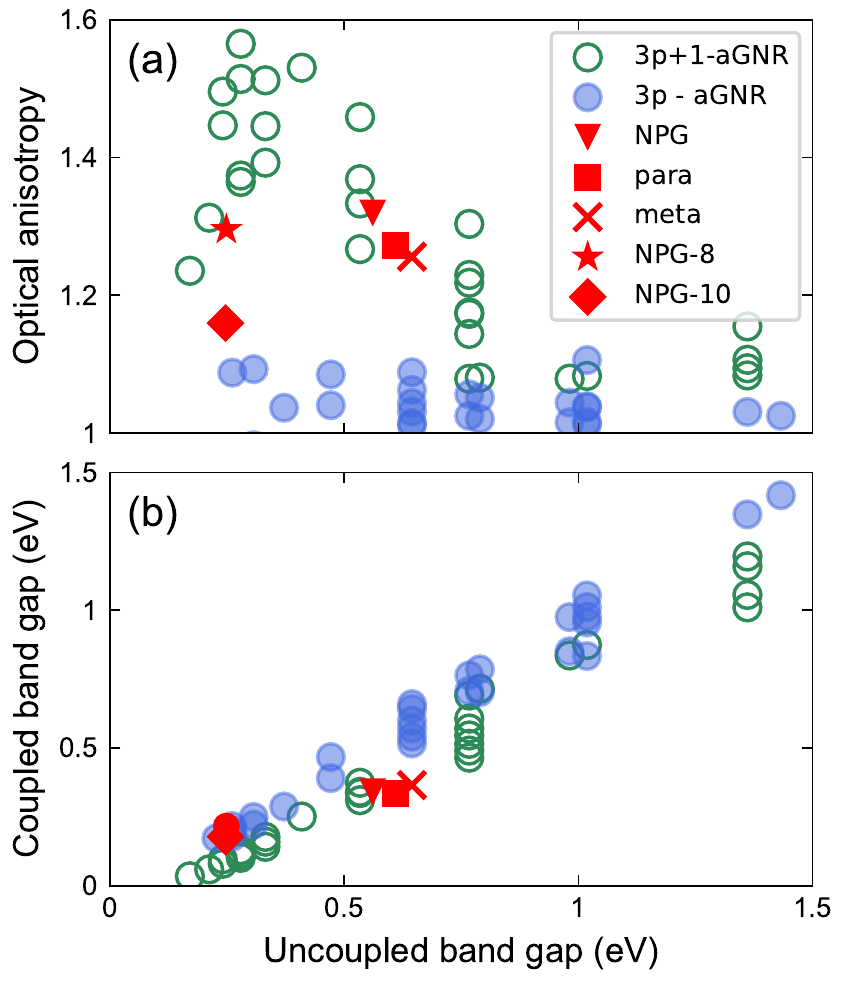}
\caption{Summary of the properties of graphene interfaced with NPG or with GNR arrays. (a) The optical anisotropy induced in graphene by a wide range of GNRs and NPGs, as a function of their band gap. (b) The energy window over which only \red{graphene-like} bands are present.}
\label{fig_summary}
\end{figure}

In Fig.\ \ref{fig_summary}, we plot a summary of our simulation results. Panel (a) shows the optical anisotropy of the heterostructures as a function of the GNR or NPG band gap. Open circles correspond to graphene interfaced with $(3p+1)$-aGNRs, solid circles to $3p$-aGNRs, and the other red symbols to the NPGs. In this set of results, we have varied the ribbon width up to \red{$m=57$, corresponding to a width of 7 nm}. For many values of $m$, we have also varied the inter-GNR distance. The optical anisotropy has been averaged over the energy range for which only \red{graphene-like} bands are involved.

Here, we see a few interesting trends. First, the $3p$-aGNRs show an almost constant and small anisotropy ($\approx$1), independent of the band gap. Meanwhile, \red{an optimal ribbon width} exists for the $(3p+1)$-aGNRs, reaching an anisotropy \red{of nearly 1.6 for $m=37$, or a width of 4.6 nm}. \red{For the NPGs, all fall within the trend of the GNRs.} We note that the meta- and para-NPGs show little difference in the anisotropy they induce in graphene, despite their significantly different transport properties in isolation \cite{Gaetano2019}. \red{Finally, the NPG-8 and NPG-10 structures fall on the other side of the optimal width/band gap. These results suggest that there may exist an as-yet undiscovered NPG structure that can optimize the anisotropy induced in graphene.}

Importantly, while reducing the band gap of the nanoribbon increases the anisotropy, it also reduces the energy range in which the linearity of the graphene bands is preserved. This denotes the maximum photon energy which could excite \red{graphene-like} states, hence giving the upper limit for absorption frequency of ``anisotropic'' graphene. We present this quantity as a function of the NPG or GNR band gap in Fig.\ \ref{fig_summary}(b). As would be expected, a smaller bandgap results in a smaller energy window for optical absorption only in the graphene layer.

To summarize, we have tested the hypothesis that anisotropic semiconducting carbon materials -- such as nanoporous graphene or arrays of graphene nanoribbons -- can induce anisotropy in graphene while also maintaining its Dirac-like band structure. We have shown that this is indeed the case, with such systems inducing an anisotropy of 20\%-50\% in graphene's Fermi velocity, optical absorption, and electrical transport. This anisotropy depends on the band gap of the NPG or GNR, but can also depend strongly on the type of GNR. This opens the possibility that other types of semiconducting GNRs or NPG structures, which may be fabricated via bottom-up synthesis \cite{Cai2010, Anthony2020}, may exhibit stronger effects than what we have presented here.

\red{Here we have studied idealized systems, while real samples may exhibit wrinkles, intercalants between the layers, or chemical functionalization of the NPG pores. With respect to wrinkles and intercalants, we have simulated the impact of distance between the graphene and NPG layers, and find that anisotropy drops below 10\% for interlayer separation $>$4 $\mathrm{\AA}$. On a large scale, the overall sample anisotropy should be the average of those regions that are strongly coupled and those that show more separation between layers. Meanwhile, simulations with functionalized NPG pores reveal that hydrogen adsorbates have a small effect on anisotropy, reducing it from 30\% to 20\% on average. Strongly doping adsorbates, such as hydroxyl or fluorine, further decrease the anistropy to 10\%, suggesting that chemical functionalization may be a key process to control in these heterostructures.}

With respect to applications, a few possibilities come to mind that may warrant further exploration. By making graphene anisotropic, this may enable its use in long-wavelength \red{(mid-IR and beyond)} optical polarimetry, which has applications in, e.g., medicine, astrophysics, or scene detection \cite{Tyo2006, Gurton2012, Demos1997, Guo2004, Ghosh2011, Aitken2004, Chrysostomou2007}. Meanwhile, anisotropic charge transport may also prove to be useful in some graphene-based sensor applications. For example, sensors based on the photothermoelectric effect in graphene may benefit from anisotropic thermal conductivity by allowing electronic heat to spread only along one direction \cite{Schuler2016, Miseikis2020, Marconi2021, Cai2014, Koppens2019, Antidormi2021}. We hope that our initial results presented here will inspire further research in these directions.

\red{Finally, we would like to note that our discussion has centered on how NPG or GNRs may induce anisotropy in graphene. However, an examination of Fig.\ \ref{fig_opt_cond} indicates that such a heterostructure may be viewed in a different light, with graphene serving to extend the anisotropic optical properties of NPG to wavelengths into mid-IR and beyond.}

\begin{acknowledgments}
ICN2 is supported by the Severo Ochoa Centres of Excellence Program, which is funded by the Spanish State Research Agency (AEI, Grant No.\ SEV-2017-0706), and is funded by the CERCA Program of the Generalitat de Catalunya. This work received funding from the European Union's Horizon 2020 Research and Innovation Program under Grant Agreement No.\ 825272 (ULISSES) and from the project ``Waveguide-Integrated Mid-Infrared Graphene Detectors for Optical Gas Sensor Systems,'' Reference No.\ PCI2018-093128, funded by the Spanish Ministry of Science and Innovation - AEI.
\end{acknowledgments}

\section*{Data Availability Statement}
The data that support the findings of this study are available from the corresponding author upon reasonable request.

\bibliography{bibliography}

\begin{thebibliography}{39}%
\makeatletter
\providecommand \@ifxundefined [1]{%
 \@ifx{#1\undefined}
}%
\providecommand \@ifnum [1]{%
 \ifnum #1\expandafter \@firstoftwo
 \else \expandafter \@secondoftwo
 \fi
}%
\providecommand \@ifx [1]{%
 \ifx #1\expandafter \@firstoftwo
 \else \expandafter \@secondoftwo
 \fi
}%
\providecommand \natexlab [1]{#1}%
\providecommand \enquote  [1]{``#1''}%
\providecommand \bibnamefont  [1]{#1}%
\providecommand \bibfnamefont [1]{#1}%
\providecommand \citenamefont [1]{#1}%
\providecommand \href@noop [0]{\@secondoftwo}%
\providecommand \href [0]{\begingroup \@sanitize@url \@href}%
\providecommand \@href[1]{\@@startlink{#1}\@@href}%
\providecommand \@@href[1]{\endgroup#1\@@endlink}%
\providecommand \@sanitize@url [0]{\catcode `\\12\catcode `\$12\catcode
  `\&12\catcode `\#12\catcode `\^12\catcode `\_12\catcode `\%12\relax}%
\providecommand \@@startlink[1]{}%
\providecommand \@@endlink[0]{}%
\providecommand \url  [0]{\begingroup\@sanitize@url \@url }%
\providecommand \@url [1]{\endgroup\@href {#1}{\urlprefix }}%
\providecommand \urlprefix  [0]{URL }%
\providecommand \Eprint [0]{\href }%
\providecommand \doibase [0]{http://dx.doi.org/}%
\providecommand \selectlanguage [0]{\@gobble}%
\providecommand \bibinfo  [0]{\@secondoftwo}%
\providecommand \bibfield  [0]{\@secondoftwo}%
\providecommand \translation [1]{[#1]}%
\providecommand \BibitemOpen [0]{}%
\providecommand \bibitemStop [0]{}%
\providecommand \bibitemNoStop [0]{.\EOS\space}%
\providecommand \EOS [0]{\spacefactor3000\relax}%
\providecommand \BibitemShut  [1]{\csname bibitem#1\endcsname}%
\let\auto@bib@innerbib\@empty
\bibitem [{\citenamefont {Ferrari}\ \emph {et~al.}(2015)\citenamefont {Ferrari}
  \emph {et~al.}}]{Ferrari2015}%
  \BibitemOpen
  \bibfield  {author} {\bibinfo {author} {\bibfnamefont {A.~C.}\ \bibnamefont
  {Ferrari}} \emph {et~al.},\ }\bibfield  {title} {\enquote {\bibinfo {title}
  {{Science and technology roadmap for graphene, related two-dimensional
  crystals, and hybrid systems}},}\ }\href {\doibase 10.1039/C4NR01600A}
  {\bibfield  {journal} {\bibinfo  {journal} {Nanoscale}\ }\textbf {\bibinfo
  {volume} {7}},\ \bibinfo {pages} {4598--4810} (\bibinfo {year}
  {2015})}\BibitemShut {NoStop}%
\bibitem [{\citenamefont {Dawlaty}\ \emph {et~al.}(2008)\citenamefont
  {Dawlaty}, \citenamefont {Shivaraman}, \citenamefont {Strait}, \citenamefont
  {George}, \citenamefont {Chandrashekhar}, \citenamefont {Rana}, \citenamefont
  {Spencer}, \citenamefont {Veksler},\ and\ \citenamefont
  {Chen}}]{Dawlaty2008}%
  \BibitemOpen
  \bibfield  {author} {\bibinfo {author} {\bibfnamefont {J.~M.}\ \bibnamefont
  {Dawlaty}}, \bibinfo {author} {\bibfnamefont {S.}~\bibnamefont {Shivaraman}},
  \bibinfo {author} {\bibfnamefont {J.}~\bibnamefont {Strait}}, \bibinfo
  {author} {\bibfnamefont {P.}~\bibnamefont {George}}, \bibinfo {author}
  {\bibfnamefont {M.}~\bibnamefont {Chandrashekhar}}, \bibinfo {author}
  {\bibfnamefont {F.}~\bibnamefont {Rana}}, \bibinfo {author} {\bibfnamefont
  {M.~G.}\ \bibnamefont {Spencer}}, \bibinfo {author} {\bibfnamefont
  {D.}~\bibnamefont {Veksler}}, \ and\ \bibinfo {author} {\bibfnamefont
  {Y.}~\bibnamefont {Chen}},\ }\bibfield  {title} {\enquote {\bibinfo {title}
  {{Measurement of the optical absorption spectra of epitaxial graphene from
  terahertz to visible}},}\ }\href {\doibase 10.1063/1.2990753} {\bibfield
  {journal} {\bibinfo  {journal} {Appl. Phys. Lett.}\ }\textbf {\bibinfo
  {volume} {93}},\ \bibinfo {pages} {131905} (\bibinfo {year}
  {2008})}\BibitemShut {NoStop}%
\bibitem [{\citenamefont {Romagnoli}\ \emph {et~al.}(2018)\citenamefont
  {Romagnoli}, \citenamefont {Sorianello}, \citenamefont {Midrio},
  \citenamefont {Koppens}, \citenamefont {Huyghebaert}, \citenamefont
  {Neumaier}, \citenamefont {Galli}, \citenamefont {Templ}, \citenamefont
  {D'Errico},\ and\ \citenamefont {Ferrari}}]{Romagnoli2018}%
  \BibitemOpen
  \bibfield  {author} {\bibinfo {author} {\bibfnamefont {M.}~\bibnamefont
  {Romagnoli}}, \bibinfo {author} {\bibfnamefont {V.}~\bibnamefont
  {Sorianello}}, \bibinfo {author} {\bibfnamefont {M.}~\bibnamefont {Midrio}},
  \bibinfo {author} {\bibfnamefont {F.~H.~L.}\ \bibnamefont {Koppens}},
  \bibinfo {author} {\bibfnamefont {C.}~\bibnamefont {Huyghebaert}}, \bibinfo
  {author} {\bibfnamefont {D.}~\bibnamefont {Neumaier}}, \bibinfo {author}
  {\bibfnamefont {P.}~\bibnamefont {Galli}}, \bibinfo {author} {\bibfnamefont
  {W.}~\bibnamefont {Templ}}, \bibinfo {author} {\bibfnamefont
  {A.}~\bibnamefont {D'Errico}}, \ and\ \bibinfo {author} {\bibfnamefont
  {A.~C.}\ \bibnamefont {Ferrari}},\ }\bibfield  {title} {\enquote {\bibinfo
  {title} {{Graphene-based integrated photonics for next-generation datacom and
  telecom}},}\ }\href {\doibase 10.1038/s41578-018-0040-9} {\bibfield
  {journal} {\bibinfo  {journal} {Nat. Rev. Mater.}\ }\textbf {\bibinfo
  {volume} {3}},\ \bibinfo {pages} {392--414} (\bibinfo {year}
  {2018})}\BibitemShut {NoStop}%
\bibitem [{\citenamefont {Schuler}\ \emph {et~al.}(2016)\citenamefont
  {Schuler}, \citenamefont {Schall}, \citenamefont {Neumaier}, \citenamefont
  {Dobusch}, \citenamefont {Bethge}, \citenamefont {Schwarz}, \citenamefont
  {Krall},\ and\ \citenamefont {Mueller}}]{Schuler2016}%
  \BibitemOpen
  \bibfield  {author} {\bibinfo {author} {\bibfnamefont {S.}~\bibnamefont
  {Schuler}}, \bibinfo {author} {\bibfnamefont {D.}~\bibnamefont {Schall}},
  \bibinfo {author} {\bibfnamefont {D.}~\bibnamefont {Neumaier}}, \bibinfo
  {author} {\bibfnamefont {L.}~\bibnamefont {Dobusch}}, \bibinfo {author}
  {\bibfnamefont {O.}~\bibnamefont {Bethge}}, \bibinfo {author} {\bibfnamefont
  {B.}~\bibnamefont {Schwarz}}, \bibinfo {author} {\bibfnamefont
  {M.}~\bibnamefont {Krall}}, \ and\ \bibinfo {author} {\bibfnamefont
  {T.}~\bibnamefont {Mueller}},\ }\bibfield  {title} {\enquote {\bibinfo
  {title} {{Controlled Generation of a p-n Junction in a Waveguide Integrated
  Graphene Photodetector}},}\ }\href {\doibase 10.1021/acs.nanolett.6b03374}
  {\bibfield  {journal} {\bibinfo  {journal} {Nano Lett.}\ }\textbf {\bibinfo
  {volume} {16}},\ \bibinfo {pages} {7107--7112} (\bibinfo {year}
  {2016})}\BibitemShut {NoStop}%
\bibitem [{\citenamefont {Mi{\v s}eikis}\ \emph {et~al.}(2020)\citenamefont
  {Mi{\v s}eikis}, \citenamefont {Marconi}, \citenamefont {Giambra},
  \citenamefont {Montanaro}, \citenamefont {Martini}, \citenamefont {Fabbri},
  \citenamefont {Pezzini}, \citenamefont {Piccinini}, \citenamefont {Forti},
  \citenamefont {Terr\'{e}s}, \citenamefont {Goykhman}, \citenamefont
  {Hamidouche}, \citenamefont {Legagneux}, \citenamefont {Sorianello},
  \citenamefont {Ferrari}, \citenamefont {Koppens}, \citenamefont {Romagnoli},\
  and\ \citenamefont {Coletti}}]{Miseikis2020}%
  \BibitemOpen
  \bibfield  {author} {\bibinfo {author} {\bibfnamefont {V.}~\bibnamefont
  {Mi{\v s}eikis}}, \bibinfo {author} {\bibfnamefont {S.}~\bibnamefont
  {Marconi}}, \bibinfo {author} {\bibfnamefont {M.~A.}\ \bibnamefont
  {Giambra}}, \bibinfo {author} {\bibfnamefont {A.}~\bibnamefont {Montanaro}},
  \bibinfo {author} {\bibfnamefont {L.}~\bibnamefont {Martini}}, \bibinfo
  {author} {\bibfnamefont {F.}~\bibnamefont {Fabbri}}, \bibinfo {author}
  {\bibfnamefont {S.}~\bibnamefont {Pezzini}}, \bibinfo {author} {\bibfnamefont
  {G.}~\bibnamefont {Piccinini}}, \bibinfo {author} {\bibfnamefont
  {S.}~\bibnamefont {Forti}}, \bibinfo {author} {\bibfnamefont
  {B.}~\bibnamefont {Terr\'{e}s}}, \bibinfo {author} {\bibfnamefont
  {I.}~\bibnamefont {Goykhman}}, \bibinfo {author} {\bibfnamefont
  {L.}~\bibnamefont {Hamidouche}}, \bibinfo {author} {\bibfnamefont
  {P.}~\bibnamefont {Legagneux}}, \bibinfo {author} {\bibfnamefont
  {V.}~\bibnamefont {Sorianello}}, \bibinfo {author} {\bibfnamefont {A.~C.}\
  \bibnamefont {Ferrari}}, \bibinfo {author} {\bibfnamefont {F.~H.~L.}\
  \bibnamefont {Koppens}}, \bibinfo {author} {\bibfnamefont {M.}~\bibnamefont
  {Romagnoli}}, \ and\ \bibinfo {author} {\bibfnamefont {C.}~\bibnamefont
  {Coletti}},\ }\bibfield  {title} {\enquote {\bibinfo {title} {{Ultrafast,
  Zero-Bias, Graphene Photodetectors with Polymeric Gate Dielectric on Passive
  Photonic Waveguides}},}\ }\href {\doibase 10.1021/acsnano.0c02738} {\bibfield
   {journal} {\bibinfo  {journal} {ACS Nano}\ }\textbf {\bibinfo {volume}
  {14}},\ \bibinfo {pages} {11190--11204} (\bibinfo {year} {2020})}\BibitemShut
  {NoStop}%
\bibitem [{\citenamefont {Marconi}\ \emph {et~al.}(2021)\citenamefont
  {Marconi}, \citenamefont {Giambra}, \citenamefont {Montanaro}, \citenamefont
  {Mi{\v s}eikis}, \citenamefont {Soresi}, \citenamefont {Tirelli},
  \citenamefont {Galli}, \citenamefont {Buchali}, \citenamefont {Templ},
  \citenamefont {Coletti}, \citenamefont {Sorianello},\ and\ \citenamefont
  {Romagnoli}}]{Marconi2021}%
  \BibitemOpen
  \bibfield  {author} {\bibinfo {author} {\bibfnamefont {S.}~\bibnamefont
  {Marconi}}, \bibinfo {author} {\bibfnamefont {M.~A.}\ \bibnamefont
  {Giambra}}, \bibinfo {author} {\bibfnamefont {A.}~\bibnamefont {Montanaro}},
  \bibinfo {author} {\bibfnamefont {V.}~\bibnamefont {Mi{\v s}eikis}}, \bibinfo
  {author} {\bibfnamefont {S.}~\bibnamefont {Soresi}}, \bibinfo {author}
  {\bibfnamefont {S.}~\bibnamefont {Tirelli}}, \bibinfo {author} {\bibfnamefont
  {P.}~\bibnamefont {Galli}}, \bibinfo {author} {\bibfnamefont
  {F.}~\bibnamefont {Buchali}}, \bibinfo {author} {\bibfnamefont
  {W.}~\bibnamefont {Templ}}, \bibinfo {author} {\bibfnamefont
  {C.}~\bibnamefont {Coletti}}, \bibinfo {author} {\bibfnamefont
  {V.}~\bibnamefont {Sorianello}}, \ and\ \bibinfo {author} {\bibfnamefont
  {M.}~\bibnamefont {Romagnoli}},\ }\bibfield  {title} {\enquote {\bibinfo
  {title} {{Photo thermal effect graphene detector featuring 105 Gbit s$^{-1}$
  NRZ and 120 Gbit s$^{-1}$ PAM4 direct detection}},}\ }\href {\doibase
  10.1038/s41467-021-21137-z} {\bibfield  {journal} {\bibinfo  {journal} {Nat.
  Commun.}\ }\textbf {\bibinfo {volume} {12}},\ \bibinfo {pages} {806}
  (\bibinfo {year} {2021})}\BibitemShut {NoStop}%
\bibitem [{\citenamefont {Cai}\ \emph {et~al.}(2014)\citenamefont {Cai},
  \citenamefont {Sushkov}, \citenamefont {Suess}, \citenamefont {Jadidi},
  \citenamefont {Jenkins}, \citenamefont {Nyakiti}, \citenamefont {Myers-Ward},
  \citenamefont {Li}, \citenamefont {Yan}, \citenamefont {Gaskill},
  \citenamefont {Murphy}, \citenamefont {Drew},\ and\ \citenamefont
  {Fuhrer}}]{Cai2014}%
  \BibitemOpen
  \bibfield  {author} {\bibinfo {author} {\bibfnamefont {X.}~\bibnamefont
  {Cai}}, \bibinfo {author} {\bibfnamefont {A.~B.}\ \bibnamefont {Sushkov}},
  \bibinfo {author} {\bibfnamefont {R.~J.}\ \bibnamefont {Suess}}, \bibinfo
  {author} {\bibfnamefont {M.~M.}\ \bibnamefont {Jadidi}}, \bibinfo {author}
  {\bibfnamefont {G.~S.}\ \bibnamefont {Jenkins}}, \bibinfo {author}
  {\bibfnamefont {L.~O.}\ \bibnamefont {Nyakiti}}, \bibinfo {author}
  {\bibfnamefont {R.~L.}\ \bibnamefont {Myers-Ward}}, \bibinfo {author}
  {\bibfnamefont {S.}~\bibnamefont {Li}}, \bibinfo {author} {\bibfnamefont
  {J.}~\bibnamefont {Yan}}, \bibinfo {author} {\bibfnamefont {D.~K.}\
  \bibnamefont {Gaskill}}, \bibinfo {author} {\bibfnamefont {T.~E.}\
  \bibnamefont {Murphy}}, \bibinfo {author} {\bibfnamefont {H.~D.}\
  \bibnamefont {Drew}}, \ and\ \bibinfo {author} {\bibfnamefont {M.~S.}\
  \bibnamefont {Fuhrer}},\ }\bibfield  {title} {\enquote {\bibinfo {title}
  {{Sensitive room-temperature terahertz detection via the photothermoelectric
  effect in graphene}},}\ }\href {\doibase 10.1038/nnano.2014.182} {\bibfield
  {journal} {\bibinfo  {journal} {Nat. Nanotechnol.}\ }\textbf {\bibinfo
  {volume} {9}},\ \bibinfo {pages} {814--819} (\bibinfo {year}
  {2014})}\BibitemShut {NoStop}%
\bibitem [{\citenamefont {Castilla}\ \emph {et~al.}(2019)\citenamefont
  {Castilla}, \citenamefont {Terr\'{e}s}, \citenamefont {Autore}, \citenamefont
  {Viti}, \citenamefont {Li}, \citenamefont {Nikitin}, \citenamefont
  {Vangelidis}, \citenamefont {Watanabe}, \citenamefont {Taniguchi},
  \citenamefont {Lidorikis}, \citenamefont {Vitiello}, \citenamefont
  {Hillenbrand}, \citenamefont {Tielrooij},\ and\ \citenamefont
  {Koppens}}]{Koppens2019}%
  \BibitemOpen
  \bibfield  {author} {\bibinfo {author} {\bibfnamefont {S.}~\bibnamefont
  {Castilla}}, \bibinfo {author} {\bibfnamefont {B.}~\bibnamefont
  {Terr\'{e}s}}, \bibinfo {author} {\bibfnamefont {M.}~\bibnamefont {Autore}},
  \bibinfo {author} {\bibfnamefont {L.}~\bibnamefont {Viti}}, \bibinfo {author}
  {\bibfnamefont {J.}~\bibnamefont {Li}}, \bibinfo {author} {\bibfnamefont
  {A.~Y.}\ \bibnamefont {Nikitin}}, \bibinfo {author} {\bibfnamefont
  {I.}~\bibnamefont {Vangelidis}}, \bibinfo {author} {\bibfnamefont
  {K.}~\bibnamefont {Watanabe}}, \bibinfo {author} {\bibfnamefont
  {T.}~\bibnamefont {Taniguchi}}, \bibinfo {author} {\bibfnamefont
  {E.}~\bibnamefont {Lidorikis}}, \bibinfo {author} {\bibfnamefont {M.~S.}\
  \bibnamefont {Vitiello}}, \bibinfo {author} {\bibfnamefont {R.}~\bibnamefont
  {Hillenbrand}}, \bibinfo {author} {\bibfnamefont {K.-J.}\ \bibnamefont
  {Tielrooij}}, \ and\ \bibinfo {author} {\bibfnamefont {F.~H.~L.}\
  \bibnamefont {Koppens}},\ }\bibfield  {title} {\enquote {\bibinfo {title}
  {{Fast and Sensitive Terahertz Detection Using an Antenna-Integrated Graphene
  pn Junction}},}\ }\href {\doibase 10.1021/acs.nanolett.8b04171} {\bibfield
  {journal} {\bibinfo  {journal} {Nano Lett.}\ }\textbf {\bibinfo {volume}
  {19}},\ \bibinfo {pages} {2765--2773} (\bibinfo {year} {2019})}\BibitemShut
  {NoStop}%
\bibitem [{\citenamefont {Koppens}, \citenamefont {Chang},\ and\ \citenamefont
  {Garc\'{i}a~de Abajo}(2011)}]{Koppens2011}%
  \BibitemOpen
  \bibfield  {author} {\bibinfo {author} {\bibfnamefont {F.~H.~L.}\
  \bibnamefont {Koppens}}, \bibinfo {author} {\bibfnamefont {D.~E.}\
  \bibnamefont {Chang}}, \ and\ \bibinfo {author} {\bibfnamefont {F.~J.}\
  \bibnamefont {Garc\'{i}a~de Abajo}},\ }\bibfield  {title} {\enquote {\bibinfo
  {title} {{Graphene Plasmonics: A Platform for Strong Light–Matter
  Interactions}},}\ }\href {\doibase 10.1021/nl201771h} {\bibfield  {journal}
  {\bibinfo  {journal} {Nano Lett.}\ }\textbf {\bibinfo {volume} {11}},\
  \bibinfo {pages} {3370--3377} (\bibinfo {year} {2011})}\BibitemShut {NoStop}%
\bibitem [{\citenamefont {Grigorenko}, \citenamefont {Polini},\ and\
  \citenamefont {Novoselov}(2012)}]{Grigorenko2012}%
  \BibitemOpen
  \bibfield  {author} {\bibinfo {author} {\bibfnamefont {A.~N.}\ \bibnamefont
  {Grigorenko}}, \bibinfo {author} {\bibfnamefont {M.}~\bibnamefont {Polini}},
  \ and\ \bibinfo {author} {\bibfnamefont {K.~S.}\ \bibnamefont {Novoselov}},\
  }\bibfield  {title} {\enquote {\bibinfo {title} {{Graphene plasmonics}},}\
  }\href {\doibase 10.1038/nphoton.2012.262} {\bibfield  {journal} {\bibinfo
  {journal} {Nat. Photon.}\ }\textbf {\bibinfo {volume} {6}},\ \bibinfo {pages}
  {749--758} (\bibinfo {year} {2012})}\BibitemShut {NoStop}%
\bibitem [{\citenamefont {Cummings}\ \emph {et~al.}(2017)\citenamefont
  {Cummings}, \citenamefont {Garcia}, \citenamefont {Fabian},\ and\
  \citenamefont {Roche}}]{Cummings2017}%
  \BibitemOpen
  \bibfield  {author} {\bibinfo {author} {\bibfnamefont {A.~W.}\ \bibnamefont
  {Cummings}}, \bibinfo {author} {\bibfnamefont {J.~H.}\ \bibnamefont
  {Garcia}}, \bibinfo {author} {\bibfnamefont {J.}~\bibnamefont {Fabian}}, \
  and\ \bibinfo {author} {\bibfnamefont {S.}~\bibnamefont {Roche}},\ }\bibfield
   {title} {\enquote {\bibinfo {title} {{Giant Spin Lifetime Anisotropy in
  Graphene Induced by Proximity Effects}},}\ }\href {\doibase
  10.1103/PhysRevLett.119.206601} {\bibfield  {journal} {\bibinfo  {journal}
  {Phys. Rev. Lett.}\ }\textbf {\bibinfo {volume} {119}},\ \bibinfo {pages}
  {206601} (\bibinfo {year} {2017})}\BibitemShut {NoStop}%
\bibitem [{\citenamefont {Ben\'{i}tez}\ \emph {et~al.}(2020)\citenamefont
  {Ben\'{i}tez}, \citenamefont {Savero~Torres}, \citenamefont {Sierra},
  \citenamefont {Timmermans}, \citenamefont {Garcia}, \citenamefont {Roche},
  \citenamefont {Costache},\ and\ \citenamefont {Valenzuela}}]{Benitez2020}%
  \BibitemOpen
  \bibfield  {author} {\bibinfo {author} {\bibfnamefont {L.~A.}\ \bibnamefont
  {Ben\'{i}tez}}, \bibinfo {author} {\bibfnamefont {W.}~\bibnamefont
  {Savero~Torres}}, \bibinfo {author} {\bibfnamefont {J.~F.}\ \bibnamefont
  {Sierra}}, \bibinfo {author} {\bibfnamefont {M.}~\bibnamefont {Timmermans}},
  \bibinfo {author} {\bibfnamefont {J.~H.}\ \bibnamefont {Garcia}}, \bibinfo
  {author} {\bibfnamefont {S.}~\bibnamefont {Roche}}, \bibinfo {author}
  {\bibfnamefont {M.~V.}\ \bibnamefont {Costache}}, \ and\ \bibinfo {author}
  {\bibfnamefont {S.~O.}\ \bibnamefont {Valenzuela}},\ }\bibfield  {title}
  {\enquote {\bibinfo {title} {{Tunable room-temperature spin galvanic and spin
  Hall effects in van der Waals heterostructures}},}\ }\href {\doibase
  10.1038/s41563-019-0575-1} {\bibfield  {journal} {\bibinfo  {journal} {Nat.
  Mater.}\ }\textbf {\bibinfo {volume} {19}},\ \bibinfo {pages} {170--175}
  (\bibinfo {year} {2020})}\BibitemShut {NoStop}%
\bibitem [{\citenamefont {Forsythe}\ \emph {et~al.}(2018)\citenamefont
  {Forsythe}, \citenamefont {Zhou}, \citenamefont {Watanabe}, \citenamefont
  {Taniguchi}, \citenamefont {Pasupathy}, \citenamefont {Moon}, \citenamefont
  {Koshino}, \citenamefont {Kim},\ and\ \citenamefont {Dean}}]{Forsythe2018}%
  \BibitemOpen
  \bibfield  {author} {\bibinfo {author} {\bibfnamefont {C.}~\bibnamefont
  {Forsythe}}, \bibinfo {author} {\bibfnamefont {X.}~\bibnamefont {Zhou}},
  \bibinfo {author} {\bibfnamefont {K.}~\bibnamefont {Watanabe}}, \bibinfo
  {author} {\bibfnamefont {T.}~\bibnamefont {Taniguchi}}, \bibinfo {author}
  {\bibfnamefont {A.}~\bibnamefont {Pasupathy}}, \bibinfo {author}
  {\bibfnamefont {P.}~\bibnamefont {Moon}}, \bibinfo {author} {\bibfnamefont
  {M.}~\bibnamefont {Koshino}}, \bibinfo {author} {\bibfnamefont
  {P.}~\bibnamefont {Kim}}, \ and\ \bibinfo {author} {\bibfnamefont {C.~R.}\
  \bibnamefont {Dean}},\ }\bibfield  {title} {\enquote {\bibinfo {title} {{Band
  structure engineering of 2D materials using patterned dielectric
  superlattices}},}\ }\href {\doibase 10.1038/s41565-018-0138-7} {\bibfield
  {journal} {\bibinfo  {journal} {Nat. Nanotechnol.}\ }\textbf {\bibinfo
  {volume} {13}},\ \bibinfo {pages} {566--571} (\bibinfo {year}
  {2018})}\BibitemShut {NoStop}%
\bibitem [{\citenamefont {Li}\ \emph {et~al.}(2021)\citenamefont {Li},
  \citenamefont {Dietrich}, \citenamefont {Forsythe}, \citenamefont
  {Taniguchi}, \citenamefont {Watanabe}, \citenamefont {Moon},\ and\
  \citenamefont {Dean}}]{Li2021}%
  \BibitemOpen
  \bibfield  {author} {\bibinfo {author} {\bibfnamefont {Y.}~\bibnamefont
  {Li}}, \bibinfo {author} {\bibfnamefont {S.}~\bibnamefont {Dietrich}},
  \bibinfo {author} {\bibfnamefont {C.}~\bibnamefont {Forsythe}}, \bibinfo
  {author} {\bibfnamefont {T.}~\bibnamefont {Taniguchi}}, \bibinfo {author}
  {\bibfnamefont {K.}~\bibnamefont {Watanabe}}, \bibinfo {author}
  {\bibfnamefont {P.}~\bibnamefont {Moon}}, \ and\ \bibinfo {author}
  {\bibfnamefont {C.~R.}\ \bibnamefont {Dean}},\ }\bibfield  {title} {\enquote
  {\bibinfo {title} {{Anisotropic band flattening in graphene with
  one-dimensional superlattices}},}\ }\href {\doibase
  10.1038/s41565-021-00849-9} {\bibfield  {journal} {\bibinfo  {journal} {Nat.
  Nanotechnol.}\ }\textbf {\bibinfo {volume} {16}},\ \bibinfo {pages}
  {525--530} (\bibinfo {year} {2021})}\BibitemShut {NoStop}%
\bibitem [{\citenamefont {Yankowitz}\ \emph {et~al.}(2012)\citenamefont
  {Yankowitz}, \citenamefont {Xue}, \citenamefont {Cormode}, \citenamefont
  {Sanchez-Yamagishi}, \citenamefont {Watanabe}, \citenamefont {Taniguchi},
  \citenamefont {Jarillo-Herrero}, \citenamefont {Jacquod},\ and\ \citenamefont
  {LeRoy}}]{Yankowitz2012}%
  \BibitemOpen
  \bibfield  {author} {\bibinfo {author} {\bibfnamefont {M.}~\bibnamefont
  {Yankowitz}}, \bibinfo {author} {\bibfnamefont {J.}~\bibnamefont {Xue}},
  \bibinfo {author} {\bibfnamefont {D.}~\bibnamefont {Cormode}}, \bibinfo
  {author} {\bibfnamefont {J.~D.}\ \bibnamefont {Sanchez-Yamagishi}}, \bibinfo
  {author} {\bibfnamefont {K.}~\bibnamefont {Watanabe}}, \bibinfo {author}
  {\bibfnamefont {T.}~\bibnamefont {Taniguchi}}, \bibinfo {author}
  {\bibfnamefont {P.}~\bibnamefont {Jarillo-Herrero}}, \bibinfo {author}
  {\bibfnamefont {P.}~\bibnamefont {Jacquod}}, \ and\ \bibinfo {author}
  {\bibfnamefont {B.~J.}\ \bibnamefont {LeRoy}},\ }\bibfield  {title} {\enquote
  {\bibinfo {title} {{Emergence of superlattice Dirac points in graphene on
  hexagonal boron nitride}},}\ }\href {\doibase 10.1038/nphys2272} {\bibfield
  {journal} {\bibinfo  {journal} {Nat. Phys.}\ }\textbf {\bibinfo {volume}
  {8}},\ \bibinfo {pages} {382--386} (\bibinfo {year} {2012})}\BibitemShut
  {NoStop}%
\bibitem [{\citenamefont {Cao}\ \emph {et~al.}(2018)\citenamefont {Cao},
  \citenamefont {Fatemi}, \citenamefont {Fang}, \citenamefont {Watanabe},
  \citenamefont {Taniguchi}, \citenamefont {Kaxiras},\ and\ \citenamefont
  {Jarillo-Herrero}}]{Cao2018}%
  \BibitemOpen
  \bibfield  {author} {\bibinfo {author} {\bibfnamefont {Y.}~\bibnamefont
  {Cao}}, \bibinfo {author} {\bibfnamefont {V.}~\bibnamefont {Fatemi}},
  \bibinfo {author} {\bibfnamefont {S.}~\bibnamefont {Fang}}, \bibinfo {author}
  {\bibfnamefont {K.}~\bibnamefont {Watanabe}}, \bibinfo {author}
  {\bibfnamefont {T.}~\bibnamefont {Taniguchi}}, \bibinfo {author}
  {\bibfnamefont {E.}~\bibnamefont {Kaxiras}}, \ and\ \bibinfo {author}
  {\bibfnamefont {P.}~\bibnamefont {Jarillo-Herrero}},\ }\bibfield  {title}
  {\enquote {\bibinfo {title} {{Unconventional superconductivity in magic-angle
  graphene superlattices}},}\ }\href {\doibase 10.1038/nature26160} {\bibfield
  {journal} {\bibinfo  {journal} {Nature}\ }\textbf {\bibinfo {volume} {556}},\
  \bibinfo {pages} {43--50} (\bibinfo {year} {2018})}\BibitemShut {NoStop}%
\bibitem [{\citenamefont {Moreno}\ \emph
  {et~al.}(2018{\natexlab{a}})\citenamefont {Moreno}, \citenamefont
  {Vilas-Varela}, \citenamefont {Kretz}, \citenamefont {Garcia-Lekue},
  \citenamefont {Costache}, \citenamefont {Paradinas}, \citenamefont
  {Panighel}, \citenamefont {Ceballos}, \citenamefont {Valenzuela},
  \citenamefont {Pe{\~n}a},\ and\ \citenamefont {Mugarza}}]{Moreno2018}%
  \BibitemOpen
  \bibfield  {author} {\bibinfo {author} {\bibfnamefont {C.}~\bibnamefont
  {Moreno}}, \bibinfo {author} {\bibfnamefont {M.}~\bibnamefont
  {Vilas-Varela}}, \bibinfo {author} {\bibfnamefont {B.}~\bibnamefont {Kretz}},
  \bibinfo {author} {\bibfnamefont {A.}~\bibnamefont {Garcia-Lekue}}, \bibinfo
  {author} {\bibfnamefont {M.~V.}\ \bibnamefont {Costache}}, \bibinfo {author}
  {\bibfnamefont {M.}~\bibnamefont {Paradinas}}, \bibinfo {author}
  {\bibfnamefont {M.}~\bibnamefont {Panighel}}, \bibinfo {author}
  {\bibfnamefont {G.}~\bibnamefont {Ceballos}}, \bibinfo {author}
  {\bibfnamefont {S.~O.}\ \bibnamefont {Valenzuela}}, \bibinfo {author}
  {\bibfnamefont {D.}~\bibnamefont {Pe{\~n}a}}, \ and\ \bibinfo {author}
  {\bibfnamefont {A.}~\bibnamefont {Mugarza}},\ }\bibfield  {title} {\enquote
  {\bibinfo {title} {{Bottom-up synthesis of multifunctional nanoporous
  graphene}},}\ }\href {\doibase 10.1126/science.aar2009} {\bibfield  {journal}
  {\bibinfo  {journal} {Science}\ }\textbf {\bibinfo {volume} {360}},\ \bibinfo
  {pages} {199--203} (\bibinfo {year} {2018}{\natexlab{a}})}\BibitemShut
  {NoStop}%
\bibitem [{\citenamefont {Moreno}\ \emph
  {et~al.}(2018{\natexlab{b}})\citenamefont {Moreno}, \citenamefont
  {Paradinas}, \citenamefont {Vilas-Varela}, \citenamefont {Panighel},
  \citenamefont {Ceballos}, \citenamefont {Pe\~{n}a},\ and\ \citenamefont
  {Mugarza}}]{Moreno2018b}%
  \BibitemOpen
  \bibfield  {author} {\bibinfo {author} {\bibfnamefont {C.}~\bibnamefont
  {Moreno}}, \bibinfo {author} {\bibfnamefont {M.}~\bibnamefont {Paradinas}},
  \bibinfo {author} {\bibfnamefont {M.}~\bibnamefont {Vilas-Varela}}, \bibinfo
  {author} {\bibfnamefont {M.}~\bibnamefont {Panighel}}, \bibinfo {author}
  {\bibfnamefont {G.}~\bibnamefont {Ceballos}}, \bibinfo {author}
  {\bibfnamefont {D.}~\bibnamefont {Pe\~{n}a}}, \ and\ \bibinfo {author}
  {\bibfnamefont {A.}~\bibnamefont {Mugarza}},\ }\bibfield  {title} {\enquote
  {\bibinfo {title} {On-surface synthesis of superlattice arrays of ultra-long
  graphene nanoribbons},}\ }\href {\doibase 10.1039/C8CC04830D} {\bibfield
  {journal} {\bibinfo  {journal} {Chem. Commun.}\ }\textbf {\bibinfo {volume}
  {54}},\ \bibinfo {pages} {9402--9405} (\bibinfo {year}
  {2018}{\natexlab{b}})}\BibitemShut {NoStop}%
\bibitem [{\citenamefont {Tyo}\ \emph {et~al.}(2006)\citenamefont {Tyo},
  \citenamefont {Goldstein}, \citenamefont {Chenault},\ and\ \citenamefont
  {Shaw}}]{Tyo2006}%
  \BibitemOpen
  \bibfield  {author} {\bibinfo {author} {\bibfnamefont {J.~S.}\ \bibnamefont
  {Tyo}}, \bibinfo {author} {\bibfnamefont {D.~L.}\ \bibnamefont {Goldstein}},
  \bibinfo {author} {\bibfnamefont {D.~B.}\ \bibnamefont {Chenault}}, \ and\
  \bibinfo {author} {\bibfnamefont {J.~A.}\ \bibnamefont {Shaw}},\ }\bibfield
  {title} {\enquote {\bibinfo {title} {{Review of passive imaging polarimetry
  for remote sensing applications}},}\ }\href {\doibase 10.1364/AO.45.005453}
  {\bibfield  {journal} {\bibinfo  {journal} {Appl. Opt.}\ }\textbf {\bibinfo
  {volume} {45}},\ \bibinfo {pages} {5453--5469} (\bibinfo {year}
  {2006})}\BibitemShut {NoStop}%
\bibitem [{\citenamefont {Gurton}\ and\ \citenamefont
  {Felton}(2012)}]{Gurton2012}%
  \BibitemOpen
  \bibfield  {author} {\bibinfo {author} {\bibfnamefont {K.~P.}\ \bibnamefont
  {Gurton}}\ and\ \bibinfo {author} {\bibfnamefont {M.}~\bibnamefont
  {Felton}},\ }\bibfield  {title} {\enquote {\bibinfo {title} {{Remote
  detection of buried land-mines and IEDs using LWIR polarimetric imaging}},}\
  }\href {\doibase 10.1364/OE.20.022344} {\bibfield  {journal} {\bibinfo
  {journal} {Opt. Express}\ }\textbf {\bibinfo {volume} {20}},\ \bibinfo
  {pages} {22344--22359} (\bibinfo {year} {2012})}\BibitemShut {NoStop}%
\bibitem [{\citenamefont {Demos}\ and\ \citenamefont
  {Alfano}(1997)}]{Demos1997}%
  \BibitemOpen
  \bibfield  {author} {\bibinfo {author} {\bibfnamefont {S.~G.}\ \bibnamefont
  {Demos}}\ and\ \bibinfo {author} {\bibfnamefont {R.~R.}\ \bibnamefont
  {Alfano}},\ }\bibfield  {title} {\enquote {\bibinfo {title} {{Optical
  polarization imaging}},}\ }\href {\doibase 10.1364/AO.36.000150} {\bibfield
  {journal} {\bibinfo  {journal} {Appl. Opt.}\ }\textbf {\bibinfo {volume}
  {36}},\ \bibinfo {pages} {150--155} (\bibinfo {year} {1997})}\BibitemShut
  {NoStop}%
\bibitem [{\citenamefont {Guo}\ \emph {et~al.}(2004)\citenamefont {Guo},
  \citenamefont {Wang}, \citenamefont {Peng}, \citenamefont {Zhang},
  \citenamefont {Luo}, \citenamefont {Le}, \citenamefont {Gmachl},
  \citenamefont {Sivco}, \citenamefont {Peabody},\ and\ \citenamefont
  {Cho}}]{Guo2004}%
  \BibitemOpen
  \bibfield  {author} {\bibinfo {author} {\bibfnamefont {B.}~\bibnamefont
  {Guo}}, \bibinfo {author} {\bibfnamefont {Y.}~\bibnamefont {Wang}}, \bibinfo
  {author} {\bibfnamefont {C.}~\bibnamefont {Peng}}, \bibinfo {author}
  {\bibfnamefont {H.~L.}\ \bibnamefont {Zhang}}, \bibinfo {author}
  {\bibfnamefont {G.~P.}\ \bibnamefont {Luo}}, \bibinfo {author} {\bibfnamefont
  {H.~Q.}\ \bibnamefont {Le}}, \bibinfo {author} {\bibfnamefont
  {C.}~\bibnamefont {Gmachl}}, \bibinfo {author} {\bibfnamefont {D.~L.}\
  \bibnamefont {Sivco}}, \bibinfo {author} {\bibfnamefont {M.~L.}\ \bibnamefont
  {Peabody}}, \ and\ \bibinfo {author} {\bibfnamefont {A.~Y.}\ \bibnamefont
  {Cho}},\ }\bibfield  {title} {\enquote {\bibinfo {title} {{Laser-based
  mid-infrared reflectance imaging of biological tissues}},}\ }\href {\doibase
  10.1364/OPEX.12.000208} {\bibfield  {journal} {\bibinfo  {journal} {Opt.
  Express}\ }\textbf {\bibinfo {volume} {12}},\ \bibinfo {pages} {208--219}
  (\bibinfo {year} {2004})}\BibitemShut {NoStop}%
\bibitem [{\citenamefont {Ghosh}\ and\ \citenamefont
  {Vitkin}(2011)}]{Ghosh2011}%
  \BibitemOpen
  \bibfield  {author} {\bibinfo {author} {\bibfnamefont {N.}~\bibnamefont
  {Ghosh}}\ and\ \bibinfo {author} {\bibfnamefont {A.~I.}\ \bibnamefont
  {Vitkin}},\ }\bibfield  {title} {\enquote {\bibinfo {title} {{Tissue
  polarimetry: concepts, challenges, applications, and outlook}},}\ }\href
  {\doibase 10.1117/1.3652896} {\bibfield  {journal} {\bibinfo  {journal} {J.
  Biomed. Opt.}\ }\textbf {\bibinfo {volume} {16}},\ \bibinfo {pages} {110801}
  (\bibinfo {year} {2011})}\BibitemShut {NoStop}%
\bibitem [{\citenamefont {Aitken}\ \emph {et~al.}(2004)\citenamefont {Aitken},
  \citenamefont {Hough}, \citenamefont {Roche}, \citenamefont {Smith},\ and\
  \citenamefont {Wright}}]{Aitken2004}%
  \BibitemOpen
  \bibfield  {author} {\bibinfo {author} {\bibfnamefont {D.~K.}\ \bibnamefont
  {Aitken}}, \bibinfo {author} {\bibfnamefont {J.~H.}\ \bibnamefont {Hough}},
  \bibinfo {author} {\bibfnamefont {P.~F.}\ \bibnamefont {Roche}}, \bibinfo
  {author} {\bibfnamefont {C.~H.}\ \bibnamefont {Smith}}, \ and\ \bibinfo
  {author} {\bibfnamefont {C.~M.}\ \bibnamefont {Wright}},\ }\bibfield  {title}
  {\enquote {\bibinfo {title} {{Mid-infrared polarimetry and magnetic fields:
  an observing strategy}},}\ }\href {\doibase 10.1111/j.1365-2966.2004.07365.x}
  {\bibfield  {journal} {\bibinfo  {journal} {Mon. Not. R. Astron. Soc.}\
  }\textbf {\bibinfo {volume} {348}},\ \bibinfo {pages} {279--284} (\bibinfo
  {year} {2004})}\BibitemShut {NoStop}%
\bibitem [{\citenamefont {Chrysostomou}, \citenamefont {Lucas},\ and\
  \citenamefont {Hough}(2007)}]{Chrysostomou2007}%
  \BibitemOpen
  \bibfield  {author} {\bibinfo {author} {\bibfnamefont {A.}~\bibnamefont
  {Chrysostomou}}, \bibinfo {author} {\bibfnamefont {P.~W.}\ \bibnamefont
  {Lucas}}, \ and\ \bibinfo {author} {\bibfnamefont {J.~H.}\ \bibnamefont
  {Hough}},\ }\bibfield  {title} {\enquote {\bibinfo {title} {{Circular
  polarimetry reveals helical magnetic fields in the young stellar object
  HH{\thinspace}135--136}},}\ }\href {\doibase 10.1038/nature06220} {\bibfield
  {journal} {\bibinfo  {journal} {Nature}\ }\textbf {\bibinfo {volume} {450}},\
  \bibinfo {pages} {71--73} (\bibinfo {year} {2007})}\BibitemShut {NoStop}%
\bibitem [{\citenamefont {Antidormi}\ and\ \citenamefont
  {Cummings}(2021)}]{Antidormi2021}%
  \BibitemOpen
  \bibfield  {author} {\bibinfo {author} {\bibfnamefont {A.}~\bibnamefont
  {Antidormi}}\ and\ \bibinfo {author} {\bibfnamefont {A.~W.}\ \bibnamefont
  {Cummings}},\ }\bibfield  {title} {\enquote {\bibinfo {title} {{Optimizing
  the Photothermoelectric Effect in Graphene}},}\ }\href {\doibase
  10.1103/PhysRevApplied.15.054049} {\bibfield  {journal} {\bibinfo  {journal}
  {Phys. Rev. Appl.}\ }\textbf {\bibinfo {volume} {15}},\ \bibinfo {pages}
  {054049} (\bibinfo {year} {2021})}\BibitemShut {NoStop}%
\bibitem [{\citenamefont {Castro~Neto}\ \emph {et~al.}(2009)\citenamefont
  {Castro~Neto}, \citenamefont {Guinea}, \citenamefont {Peres}, \citenamefont
  {Novoselov},\ and\ \citenamefont {Geim}}]{CastroNeto2009}%
  \BibitemOpen
  \bibfield  {author} {\bibinfo {author} {\bibfnamefont {A.~H.}\ \bibnamefont
  {Castro~Neto}}, \bibinfo {author} {\bibfnamefont {F.}~\bibnamefont {Guinea}},
  \bibinfo {author} {\bibfnamefont {N.~M.~R.}\ \bibnamefont {Peres}}, \bibinfo
  {author} {\bibfnamefont {K.~S.}\ \bibnamefont {Novoselov}}, \ and\ \bibinfo
  {author} {\bibfnamefont {A.~K.}\ \bibnamefont {Geim}},\ }\bibfield  {title}
  {\enquote {\bibinfo {title} {{The electronic properties of graphene}},}\
  }\href {\doibase 10.1103/RevModPhys.81.109} {\bibfield  {journal} {\bibinfo
  {journal} {Rev. Mod. Phys.}\ }\textbf {\bibinfo {volume} {81}},\ \bibinfo
  {pages} {109--162} (\bibinfo {year} {2009})}\BibitemShut {NoStop}%
\bibitem [{\citenamefont {Davies}(1997)}]{Davies1997}%
  \BibitemOpen
  \bibfield  {author} {\bibinfo {author} {\bibfnamefont {J.~H.}\ \bibnamefont
  {Davies}},\ }\href {\doibase 10.1017/CBO9780511819070} {\emph {\bibinfo
  {title} {The Physics of Low-Dimensional Semiconductors}}}\ (\bibinfo
  {publisher} {Cambridge University Press},\ \bibinfo {year}
  {1997})\BibitemShut {NoStop}%
\bibitem [{\citenamefont {Yang}\ \emph {et~al.}(2009)\citenamefont {Yang},
  \citenamefont {Deslippe}, \citenamefont {Park}, \citenamefont {Cohen},\ and\
  \citenamefont {Louie}}]{Yang2009}%
  \BibitemOpen
  \bibfield  {author} {\bibinfo {author} {\bibfnamefont {L.}~\bibnamefont
  {Yang}}, \bibinfo {author} {\bibfnamefont {J.}~\bibnamefont {Deslippe}},
  \bibinfo {author} {\bibfnamefont {C.-H.}\ \bibnamefont {Park}}, \bibinfo
  {author} {\bibfnamefont {M.~L.}\ \bibnamefont {Cohen}}, \ and\ \bibinfo
  {author} {\bibfnamefont {S.~G.}\ \bibnamefont {Louie}},\ }\bibfield  {title}
  {\enquote {\bibinfo {title} {{Excitonic Effects on the Optical Response of
  Graphene and Bilayer Graphene}},}\ }\href {\doibase
  10.1103/PhysRevLett.103.186802} {\bibfield  {journal} {\bibinfo  {journal}
  {Phys. Rev. Lett.}\ }\textbf {\bibinfo {volume} {103}},\ \bibinfo {pages}
  {186802} (\bibinfo {year} {2009})}\BibitemShut {NoStop}%
\bibitem [{\citenamefont {Singh}, \citenamefont {Shukla},\ and\ \citenamefont
  {Ahuja}(2020)}]{Singh2020}%
  \BibitemOpen
  \bibfield  {author} {\bibinfo {author} {\bibfnamefont {D.}~\bibnamefont
  {Singh}}, \bibinfo {author} {\bibfnamefont {V.}~\bibnamefont {Shukla}}, \
  and\ \bibinfo {author} {\bibfnamefont {R.}~\bibnamefont {Ahuja}},\ }\bibfield
   {title} {\enquote {\bibinfo {title} {{Optical excitations and thermoelectric
  properties of two-dimensional holey graphene}},}\ }\href {\doibase
  10.1103/PhysRevB.102.075444} {\bibfield  {journal} {\bibinfo  {journal}
  {Phys. Rev. B}\ }\textbf {\bibinfo {volume} {102}},\ \bibinfo {pages}
  {075444} (\bibinfo {year} {2020})}\BibitemShut {NoStop}%
\bibitem [{\citenamefont {Gusynin}, \citenamefont {Sharapov},\ and\
  \citenamefont {Carbotte}(2006)}]{Gusynin2006}%
  \BibitemOpen
  \bibfield  {author} {\bibinfo {author} {\bibfnamefont {V.~P.}\ \bibnamefont
  {Gusynin}}, \bibinfo {author} {\bibfnamefont {S.~G.}\ \bibnamefont
  {Sharapov}}, \ and\ \bibinfo {author} {\bibfnamefont {J.~P.}\ \bibnamefont
  {Carbotte}},\ }\bibfield  {title} {\enquote {\bibinfo {title} {{Unusual
  Microwave Response of Dirac Quasiparticles in Graphene}},}\ }\href {\doibase
  10.1103/PhysRevLett.96.256802} {\bibfield  {journal} {\bibinfo  {journal}
  {Phys. Rev. Lett.}\ }\textbf {\bibinfo {volume} {96}},\ \bibinfo {pages}
  {256802} (\bibinfo {year} {2006})}\BibitemShut {NoStop}%
\bibitem [{\citenamefont {He}\ \emph {et~al.}(2013)\citenamefont {He},
  \citenamefont {Wang}, \citenamefont {Nanot}, \citenamefont {Cong},
  \citenamefont {Jiang}, \citenamefont {Kane}, \citenamefont {Goldsmith},
  \citenamefont {Hauge}, \citenamefont {L{\'e}onard},\ and\ \citenamefont
  {Kono}}]{He2013}%
  \BibitemOpen
  \bibfield  {author} {\bibinfo {author} {\bibfnamefont {X.}~\bibnamefont
  {He}}, \bibinfo {author} {\bibfnamefont {X.}~\bibnamefont {Wang}}, \bibinfo
  {author} {\bibfnamefont {S.}~\bibnamefont {Nanot}}, \bibinfo {author}
  {\bibfnamefont {K.}~\bibnamefont {Cong}}, \bibinfo {author} {\bibfnamefont
  {Q.}~\bibnamefont {Jiang}}, \bibinfo {author} {\bibfnamefont {A.~A.}\
  \bibnamefont {Kane}}, \bibinfo {author} {\bibfnamefont {J.~E.~M.}\
  \bibnamefont {Goldsmith}}, \bibinfo {author} {\bibfnamefont {R.~H.}\
  \bibnamefont {Hauge}}, \bibinfo {author} {\bibfnamefont {F.}~\bibnamefont
  {L{\'e}onard}}, \ and\ \bibinfo {author} {\bibfnamefont {J.}~\bibnamefont
  {Kono}},\ }\bibfield  {title} {\enquote {\bibinfo {title}
  {{Photothermoelectric p-n Junction Photodetector with Intrinsic Broadband
  Polarimetry Based on Macroscopic Carbon Nanotube Films}},}\ }\href {\doibase
  10.1021/nn402679u} {\bibfield  {journal} {\bibinfo  {journal} {ACS Nano}\
  }\textbf {\bibinfo {volume} {7}},\ \bibinfo {pages} {7271--7277} (\bibinfo
  {year} {2013})}\BibitemShut {NoStop}%
\bibitem [{\citenamefont {Fan}\ \emph {et~al.}(2021)\citenamefont {Fan},
  \citenamefont {Garcia}, \citenamefont {Cummings}, \citenamefont
  {Barrios-Vargas}, \citenamefont {Panhans}, \citenamefont {Harju},
  \citenamefont {Ortmann},\ and\ \citenamefont {Roche}}]{Fan2020}%
  \BibitemOpen
  \bibfield  {author} {\bibinfo {author} {\bibfnamefont {Z.}~\bibnamefont
  {Fan}}, \bibinfo {author} {\bibfnamefont {J.~H.}\ \bibnamefont {Garcia}},
  \bibinfo {author} {\bibfnamefont {A.~W.}\ \bibnamefont {Cummings}}, \bibinfo
  {author} {\bibfnamefont {J.~E.}\ \bibnamefont {Barrios-Vargas}}, \bibinfo
  {author} {\bibfnamefont {M.}~\bibnamefont {Panhans}}, \bibinfo {author}
  {\bibfnamefont {A.}~\bibnamefont {Harju}}, \bibinfo {author} {\bibfnamefont
  {F.}~\bibnamefont {Ortmann}}, \ and\ \bibinfo {author} {\bibfnamefont
  {S.}~\bibnamefont {Roche}},\ }\bibfield  {title} {\enquote {\bibinfo {title}
  {{Linear scaling quantum transport methodologies}},}\ }\href {\doibase
  10.1016/j.physrep.2020.12.001} {\bibfield  {journal} {\bibinfo  {journal}
  {Phys. Rep.}\ }\textbf {\bibinfo {volume} {903}},\ \bibinfo {pages} {1--69}
  (\bibinfo {year} {2021})}\BibitemShut {NoStop}%
\bibitem [{\citenamefont {Adam}, \citenamefont {Brouwer},\ and\ \citenamefont
  {Das~Sarma}(2009)}]{Adam2009}%
  \BibitemOpen
  \bibfield  {author} {\bibinfo {author} {\bibfnamefont {S.}~\bibnamefont
  {Adam}}, \bibinfo {author} {\bibfnamefont {P.~W.}\ \bibnamefont {Brouwer}}, \
  and\ \bibinfo {author} {\bibfnamefont {S.}~\bibnamefont {Das~Sarma}},\
  }\bibfield  {title} {\enquote {\bibinfo {title} {{Crossover from quantum to
  Boltzmann transport in graphene}},}\ }\href {\doibase
  10.1103/PhysRevB.79.201404} {\bibfield  {journal} {\bibinfo  {journal} {Phys.
  Rev. B}\ }\textbf {\bibinfo {volume} {79}},\ \bibinfo {pages} {201404}
  (\bibinfo {year} {2009})}\BibitemShut {NoStop}%
\bibitem [{\citenamefont {Chen}\ \emph {et~al.}(2008)\citenamefont {Chen},
  \citenamefont {Jang}, \citenamefont {Adam}, \citenamefont {Fuhrer},
  \citenamefont {Williams},\ and\ \citenamefont {Ishigami}}]{Chen2008}%
  \BibitemOpen
  \bibfield  {author} {\bibinfo {author} {\bibfnamefont {J.-H.}\ \bibnamefont
  {Chen}}, \bibinfo {author} {\bibfnamefont {C.}~\bibnamefont {Jang}}, \bibinfo
  {author} {\bibfnamefont {S.}~\bibnamefont {Adam}}, \bibinfo {author}
  {\bibfnamefont {M.~S.}\ \bibnamefont {Fuhrer}}, \bibinfo {author}
  {\bibfnamefont {E.~D.}\ \bibnamefont {Williams}}, \ and\ \bibinfo {author}
  {\bibfnamefont {M.}~\bibnamefont {Ishigami}},\ }\bibfield  {title} {\enquote
  {\bibinfo {title} {{Charged-impurity scattering in graphene}},}\ }\href
  {\doibase 10.1038/nphys935} {\bibfield  {journal} {\bibinfo  {journal} {Nat.
  Phys.}\ }\textbf {\bibinfo {volume} {4}},\ \bibinfo {pages} {377--381}
  (\bibinfo {year} {2008})}\BibitemShut {NoStop}%
\bibitem [{\citenamefont {Calogero}\ \emph {et~al.}(2019)\citenamefont
  {Calogero}, \citenamefont {Alc{\'o}n}, \citenamefont {Papior}, \citenamefont
  {Jauho},\ and\ \citenamefont {Brandbyge}}]{Gaetano2019}%
  \BibitemOpen
  \bibfield  {author} {\bibinfo {author} {\bibfnamefont {G.}~\bibnamefont
  {Calogero}}, \bibinfo {author} {\bibfnamefont {I.}~\bibnamefont {Alc{\'o}n}},
  \bibinfo {author} {\bibfnamefont {N.}~\bibnamefont {Papior}}, \bibinfo
  {author} {\bibfnamefont {A.-P.}\ \bibnamefont {Jauho}}, \ and\ \bibinfo
  {author} {\bibfnamefont {M.}~\bibnamefont {Brandbyge}},\ }\bibfield  {title}
  {\enquote {\bibinfo {title} {{Quantum Interference Engineering of Nanoporous
  Graphene for Carbon Nanocircuitry}},}\ }\href {\doibase 10.1021/jacs.9b04649}
  {\bibfield  {journal} {\bibinfo  {journal} {J. Am. Chem. Soc.}\ }\textbf
  {\bibinfo {volume} {141}},\ \bibinfo {pages} {13081--13088} (\bibinfo {year}
  {2019})}\BibitemShut {NoStop}%
\bibitem [{\citenamefont {Nakada}\ \emph {et~al.}(1996)\citenamefont {Nakada},
  \citenamefont {Fujita}, \citenamefont {Dresselhaus},\ and\ \citenamefont
  {Dresselhaus}}]{Nakada1996}%
  \BibitemOpen
  \bibfield  {author} {\bibinfo {author} {\bibfnamefont {K.}~\bibnamefont
  {Nakada}}, \bibinfo {author} {\bibfnamefont {M.}~\bibnamefont {Fujita}},
  \bibinfo {author} {\bibfnamefont {G.}~\bibnamefont {Dresselhaus}}, \ and\
  \bibinfo {author} {\bibfnamefont {M.~S.}\ \bibnamefont {Dresselhaus}},\
  }\bibfield  {title} {\enquote {\bibinfo {title} {{Edge state in graphene
  ribbons: Nanometer size effect and edge shape dependence}},}\ }\href
  {\doibase 10.1103/PhysRevB.54.17954} {\bibfield  {journal} {\bibinfo
  {journal} {Phys. Rev. B}\ }\textbf {\bibinfo {volume} {54}},\ \bibinfo
  {pages} {17954--17961} (\bibinfo {year} {1996})}\BibitemShut {NoStop}%
\bibitem [{\citenamefont {Cai}\ \emph {et~al.}(2010)\citenamefont {Cai},
  \citenamefont {Ruffieux}, \citenamefont {Jaafar}, \citenamefont {Bieri},
  \citenamefont {Braun}, \citenamefont {Blankenburg}, \citenamefont {Muoth},
  \citenamefont {Seitsonen}, \citenamefont {Saleh}, \citenamefont {Feng},
  \citenamefont {M{\"u}llen},\ and\ \citenamefont {Fasel}}]{Cai2010}%
  \BibitemOpen
  \bibfield  {author} {\bibinfo {author} {\bibfnamefont {J.}~\bibnamefont
  {Cai}}, \bibinfo {author} {\bibfnamefont {P.}~\bibnamefont {Ruffieux}},
  \bibinfo {author} {\bibfnamefont {R.}~\bibnamefont {Jaafar}}, \bibinfo
  {author} {\bibfnamefont {M.}~\bibnamefont {Bieri}}, \bibinfo {author}
  {\bibfnamefont {T.}~\bibnamefont {Braun}}, \bibinfo {author} {\bibfnamefont
  {S.}~\bibnamefont {Blankenburg}}, \bibinfo {author} {\bibfnamefont
  {M.}~\bibnamefont {Muoth}}, \bibinfo {author} {\bibfnamefont {A.~P.}\
  \bibnamefont {Seitsonen}}, \bibinfo {author} {\bibfnamefont {M.}~\bibnamefont
  {Saleh}}, \bibinfo {author} {\bibfnamefont {X.}~\bibnamefont {Feng}},
  \bibinfo {author} {\bibfnamefont {K.}~\bibnamefont {M{\"u}llen}}, \ and\
  \bibinfo {author} {\bibfnamefont {R.}~\bibnamefont {Fasel}},\ }\bibfield
  {title} {\enquote {\bibinfo {title} {{Atomically precise bottom-up
  fabrication of graphene nanoribbons}},}\ }\href {\doibase
  10.1038/nature09211} {\bibfield  {journal} {\bibinfo  {journal} {Nature}\
  }\textbf {\bibinfo {volume} {466}},\ \bibinfo {pages} {470--473} (\bibinfo
  {year} {2010})}\BibitemShut {NoStop}%
\bibitem [{\citenamefont {Jolly}\ \emph {et~al.}(2020)\citenamefont {Jolly},
  \citenamefont {Miao}, \citenamefont {Daigle},\ and\ \citenamefont
  {Morin}}]{Anthony2020}%
  \BibitemOpen
  \bibfield  {author} {\bibinfo {author} {\bibfnamefont {A.}~\bibnamefont
  {Jolly}}, \bibinfo {author} {\bibfnamefont {D.}~\bibnamefont {Miao}},
  \bibinfo {author} {\bibfnamefont {M.}~\bibnamefont {Daigle}}, \ and\ \bibinfo
  {author} {\bibfnamefont {J.-F.}\ \bibnamefont {Morin}},\ }\bibfield  {title}
  {\enquote {\bibinfo {title} {{Emerging Bottom-Up Strategies for the Synthesis
  of Graphene Nanoribbons and Related Structures}},}\ }\href {\doibase
  https://doi.org/10.1002/anie.201906379} {\bibfield  {journal} {\bibinfo
  {journal} {Angew. Chem. Int. Ed.}\ }\textbf {\bibinfo {volume} {59}},\
  \bibinfo {pages} {4624--4633} (\bibinfo {year} {2020})}\BibitemShut {NoStop}%
\end{thebibliography}%
\end{document}